\documentclass[prd,eqsecnum,amsfonts,amssymb,amsmath,floatfix,nofootinbib]{revtex4}

\usepackage{bm}
\usepackage{slashed}
\usepackage{graphicx}
\usepackage{MnSymbol}

\begin{document}

\title{A dark matter WIMP that can be detected and definitively identified \\
with currently planned experiments}

\author{Caden LaFontaine, Bailey Tallman, Spencer Ellis, Trevor Croteau, Brandon Torres, Sabrina Hernandez, Diego Cristancho Guerrero, Jessica Jaksik, Drue Lubanski, and Roland Allen}

\affiliation{Department of Physics and Astronomy, 
Texas A\&M University, College Station, Texas 77843, USA}

\date{\today}

\begin{abstract}
A recently proposed dark matter WIMP has only second-order couplings to gauge bosons and itself. As a result, it has small annihilation, scattering, and creation cross-sections, and is consequently consistent with all current experiments and the observed abundance of dark matter. These cross-sections are, however, still sufficiently large to enable detection in experiments that are planned for the near future, and definitive identification in experiments proposed on a longer time scale. The (multi-channel) cross-section for annihilation is consistent with thermal production and freeze-out  in the early universe, and with current evidence for dark matter annihilation in analyses of the observations of gamma rays by Fermi-LAT and antiprotons by AMS-02, as well as the constraints from Planck and Fermi-LAT. The cross-section for direct detection via collision with xenon nuclei is estimated to be slightly below $10^{-47}$ cm$^2$, which should be attainable by LZ and Xenon nT and well within the reach of Darwin. The cross-section for collider detection via vector boson fusion is estimated to be $\sim 1$ fb, and may be ultimately  attainable by the high-luminosity LHC; definitive collider identification will probably require the more powerful facilities now being proposed.
\end{abstract}

\maketitle

\section{\label{sec:sec1}Introduction}

Following many brilliant innovations during the past third of a century, current dark matter experiments have achieved amazing sensitivities~\cite{Aprile-2018,Aprile-2019,LUX,SuperCDMS,PandaX,PICO,Icecube,Strigari-2014,Freese,2019-review,Rauch,Lisanti,Klasen,Silk,indirect-2011,Slatyer,Hooper-2018,Strigari,Planck,2018-review,Kahlhoefer,ATLAS-CMS,CMS-2019,ATLAS-2019}, imposing stringent constraints on any theoretical dark matter candidate~\cite{snowmass-2013,Baer-Roszkowski-2015,Olive-2015,Baer-Barger-2016,Baer-Barger-2018,Roszkowski-2018,Baer-Barger-2020,higgsino-2018,pdg}. In particular, the most simplistic models with supersymmetry (susy) and weakly interacting massive particles (WIMPs) have been disconfirmed by experiment, and this has led some to increasing pessimism about their existence~\cite{Bertone,Peskin-2015,Olive1}. But there are still quite compelling arguments for both susy~\cite{Haber-Kane,Baer-Tata,Kane,pdg-susy} and WIMPs~\cite{susy-DM-1996,Kamionkowski-1997}. A natural supersymmetric dark matter particle is still possible in a multicomponent dark matter scenario, containing both, e.g., the lightest neutralino of susy and some other constituent such as the axion~\cite{Kamionkowski-1997,Baer-Barger-2011,
Baer-Barger-Tata-2012,Baer-Barger-Tata-2013,Tata-2020,PQ,Weinberg,Wilczek,Sikivie,Tanner,axion-review-1,axion-review-2}.

Here, however, we will focus on a recently proposed WIMP~\cite{DM4} which does not require susy, which can account for the dark matter either by itself or as part of a multicomponent scenario, and which is fully consistent with experiment and observation. This particle is the lowest-mass ``higgson"~\cite{DM4}, represented by $H^0$ to distinguish it from the lowest mass Higgs boson $h^0$ and higgsino $\widetilde{h}^0$. 

\section{\label{sec:sec2}Mathematical formulation}

Before gauge interactions are added, the action for a generic 4-component higgson field $H$ is
\begin{align}
S_H & =\int d^{4}x\,H^{\dag }\left( x\right) \left( \partial^{\mu }\partial_{\mu}-
m^{2}\right) H\left( x\right)  \; .
\label{eq1}
\end{align}
As described in Ref.~\cite{DM4}, these and Higgs/amplitude modes are the Lorentz-invariant mass eigenstates in an extended Higgs sector. 

The gauge interactions of $H^0$ are given by~\cite{DM4}
\begin{align}
\mathcal{L}_{0}^{Z}=-\frac{g_{Z}^{2}}{4}H^{0\dag }Z^{\mu }Z_{\mu
}H^{0}\;\;,\;\;\mathcal{L}_{0}^{W}=-\frac{g^{2}}{2}H^{0\dag }W^{\mu +}W_{\mu
}^{-}H^{0}
\label{eq2}
\end{align}
where $g_{Z}=g/\cos \theta _{w}$. 
The original electroweak gauge covariant derivative after symmetry-breaking~\cite{peskin} 
\begin{align}
D_{\mu }=\partial _{\mu } -i\frac{g}{\sqrt{2}}\left( W_{\mu}^{+}T^{+}+W_{\mu }^{-}T^{-}\right)   
 -i\frac{g}{\cos \theta _{w}}Z_{\mu }\left( T^{3}-\sin ^{2}\theta
_{w}\,Q\right) -ieA_{\mu }\,Q  
\end{align}
with
\begin{align}
\partial ^{\mu }Z_{\mu }=0 \quad ,\quad \partial ^{\mu }W_{\mu }^{\pm }=0 , \quad , \quad \partial ^{\mu }A_{\mu }=0 
\end{align} 
reduces to the simple second-order interactions of (\ref{eq2}) because $H^{0}$ is one of the real components of a Majorana-like bosonic field, for which the first-order gauge interactions vanish, with the generic form
\begin{align}
\Phi_{S}= H_{S}+ i H_{S}^{\prime} \quad , \quad
\Phi_{S}=\frac{1}{\sqrt{2}}\left( 
\begin{array}{c}
\Phi _{s} \\ 
\Phi _{s}^{c}
\end{array}
\right) \;.
\end{align}
Here each 4-component complex field $\Phi _{S}$ consists of a primitive 2-component field $\Phi _{s}$ and its charge conjugate.

The present theory provides no interaction of the higgson $H^0$ with the Higgs boson $h^0$, so we will take this interaction to be zero. This implies that:

(i) The dark matter particle $H^0$ interacts with only the $Z$ and $W$ gauge bosons (and itself), and 

(ii) the role of susy in protecting the mass of the Higgs boson is unchanged, with the usual cancellation of radiative corrections between bosons and fermions. The present picture is thus compatible with susy (while not requiring it). 

\begin{figure}
\begin{center}
\resizebox{0.25\columnwidth}{!}{
\includegraphics{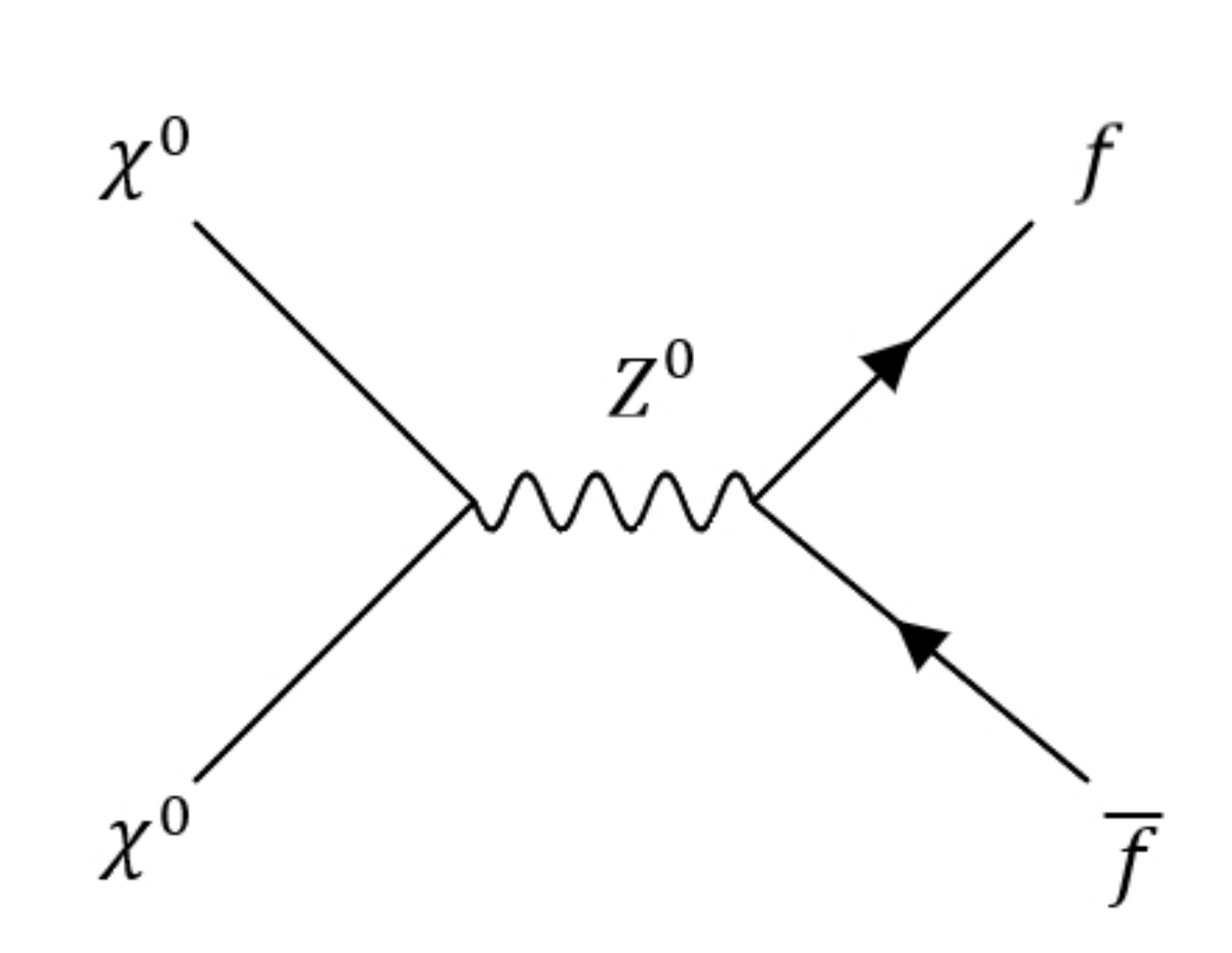}}
\resizebox{0.18\columnwidth}{!}{
\includegraphics{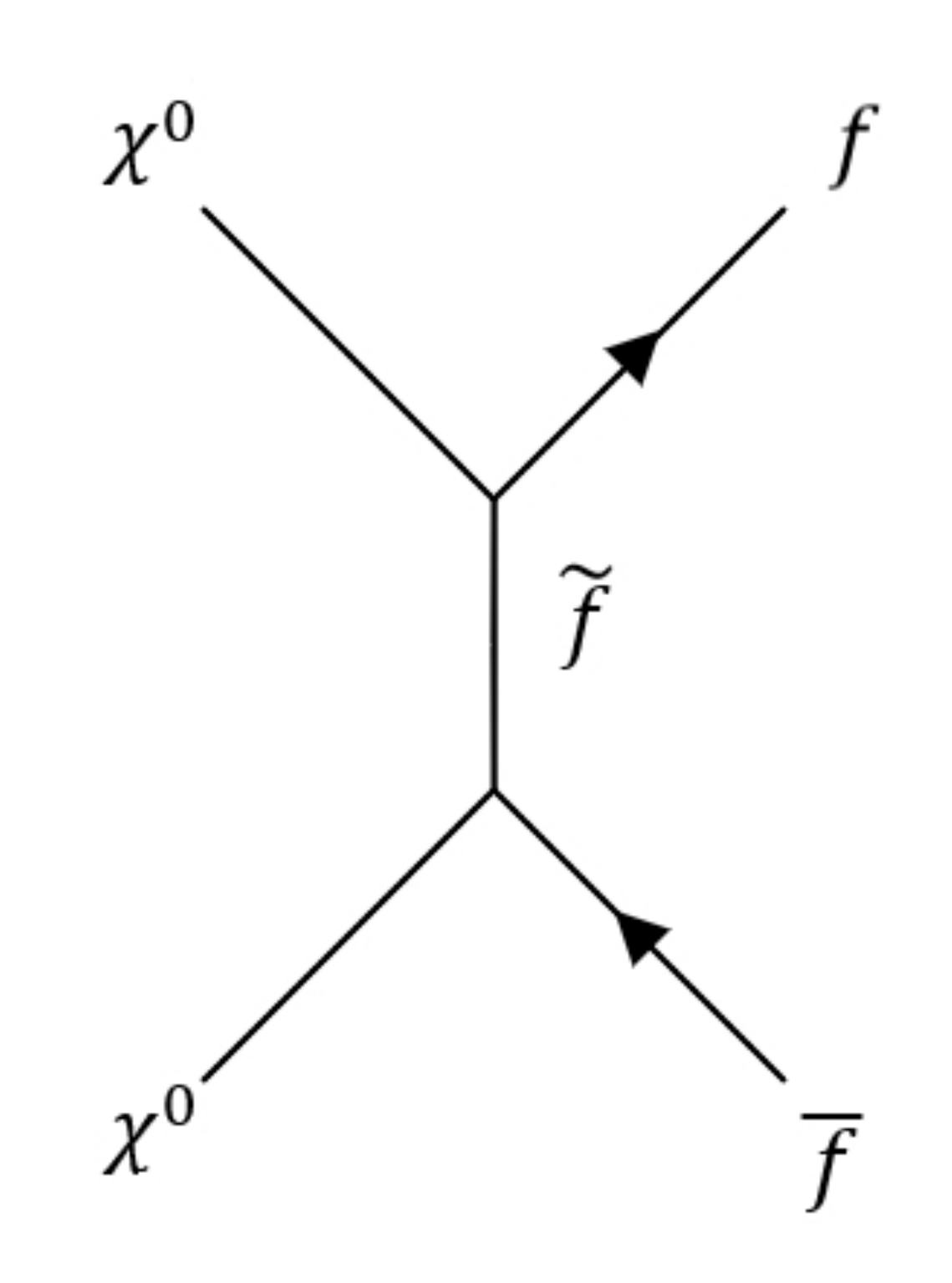}}
\resizebox{0.25\columnwidth}{!}{
\includegraphics{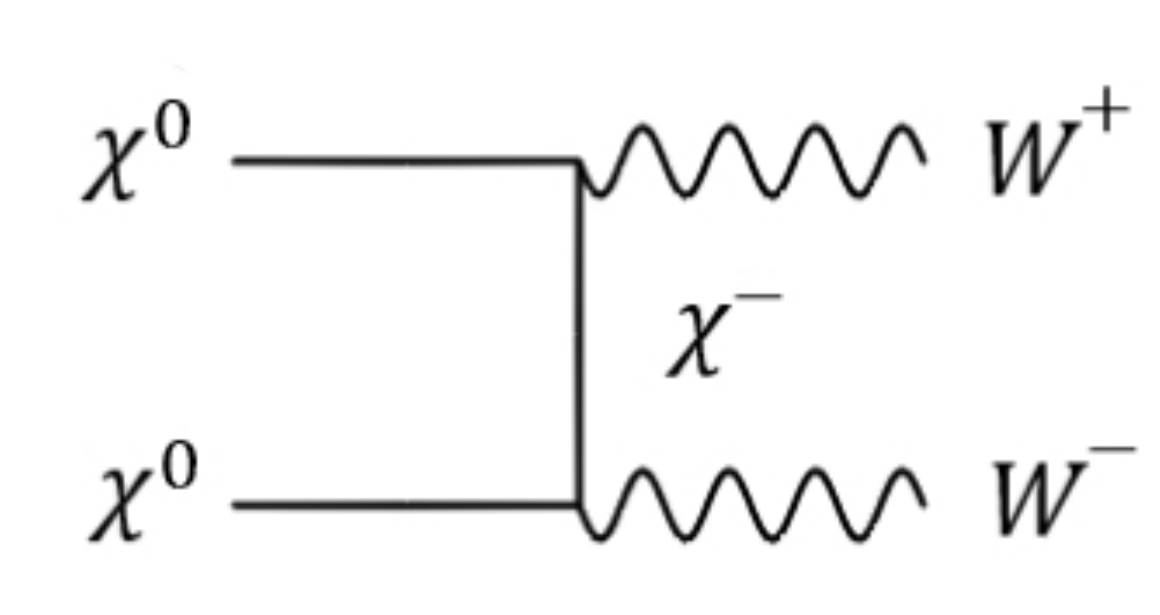}}
\resizebox{0.25\columnwidth}{!}{
\includegraphics{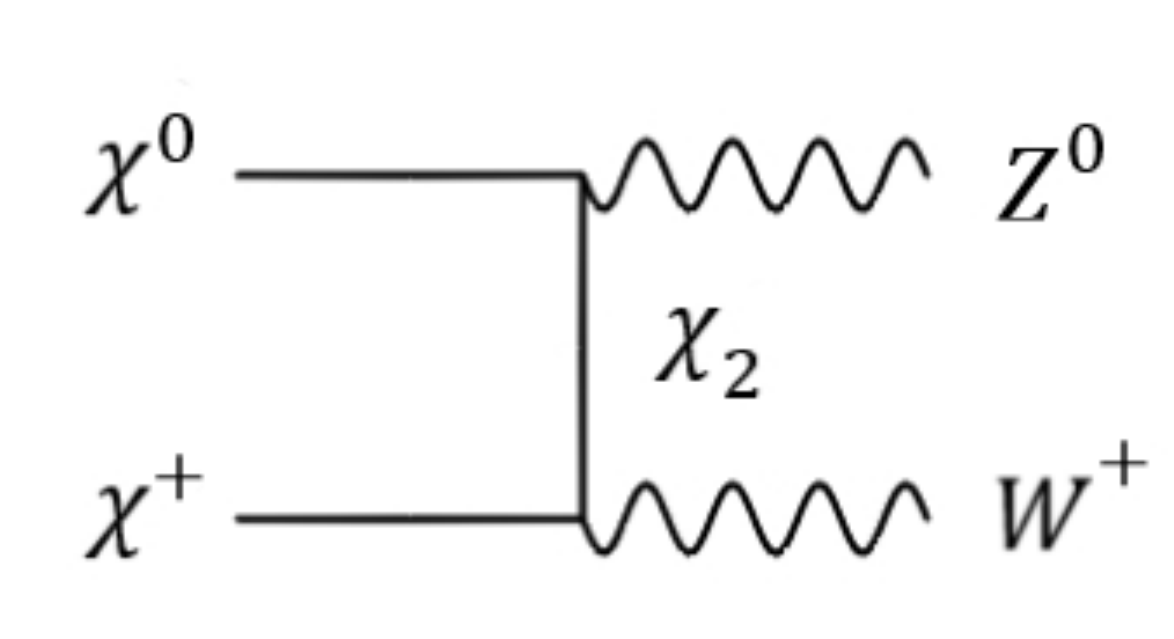}}
\end{center}
\caption{Left to right: Annihilation to fermion-antifermion pairs via $Z^0$ exchange, and sfermion exchange, occurs for neutralinos but not higgsons. (These left two figures follow Ref.~\cite{susy-DM-1996}, retaining the same conventions.) Annihilation to $W^+$ $W^-$ pairs via $\chi^{\pm}$ exchange, and coannihilation to $Z^0$ $W^+$  via $\chi_2$ exchange, occur for neutralinos but not higgons. (These right two figures follow Ref.~\cite{higgsino-2018}, retaining the same conventions.)}
\label{fig1}    
\end{figure}

\begin{figure}
\begin{center}
\resizebox{0.20\columnwidth}{!}{
\includegraphics{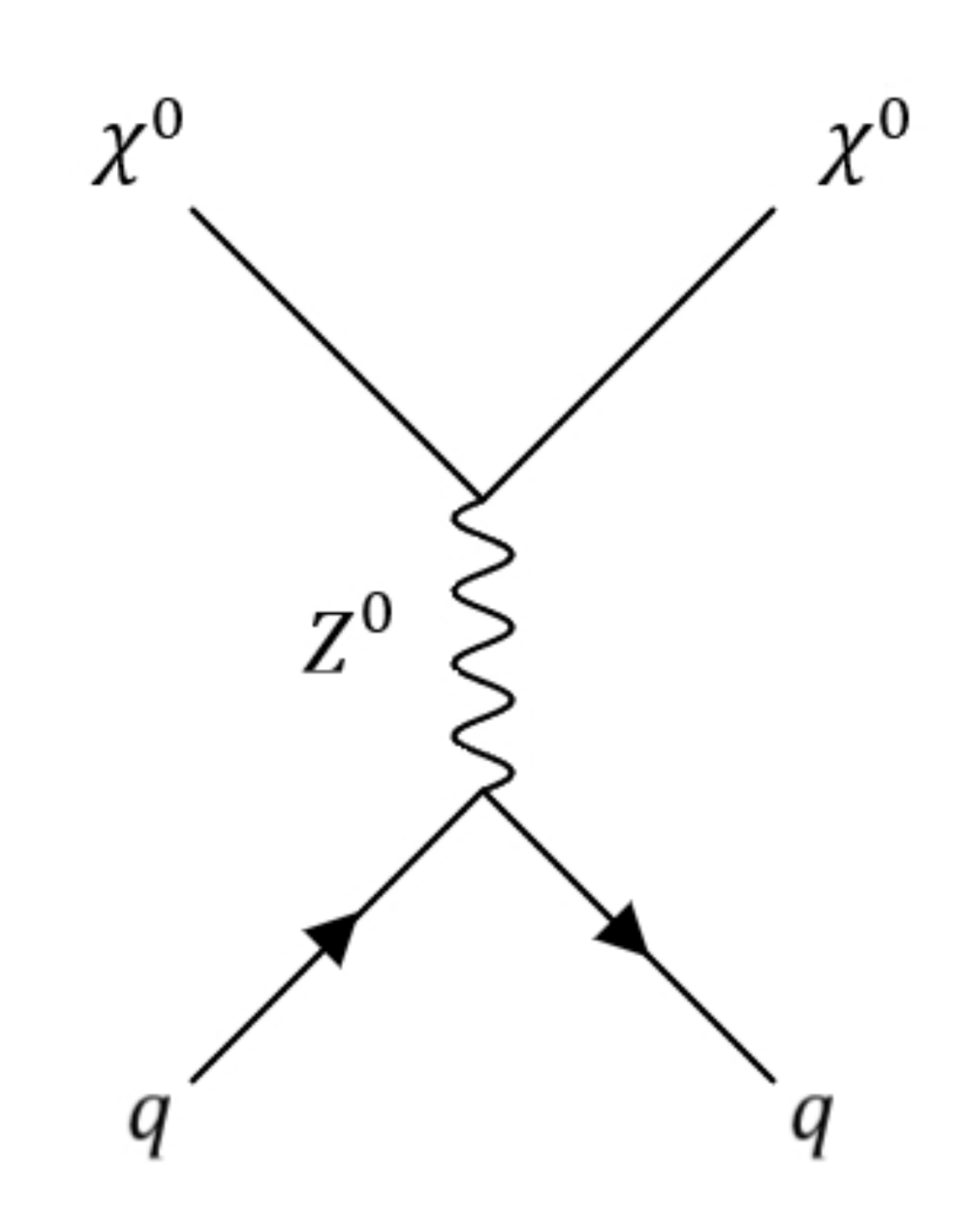}}
\resizebox{0.27\columnwidth}{!}{
\includegraphics{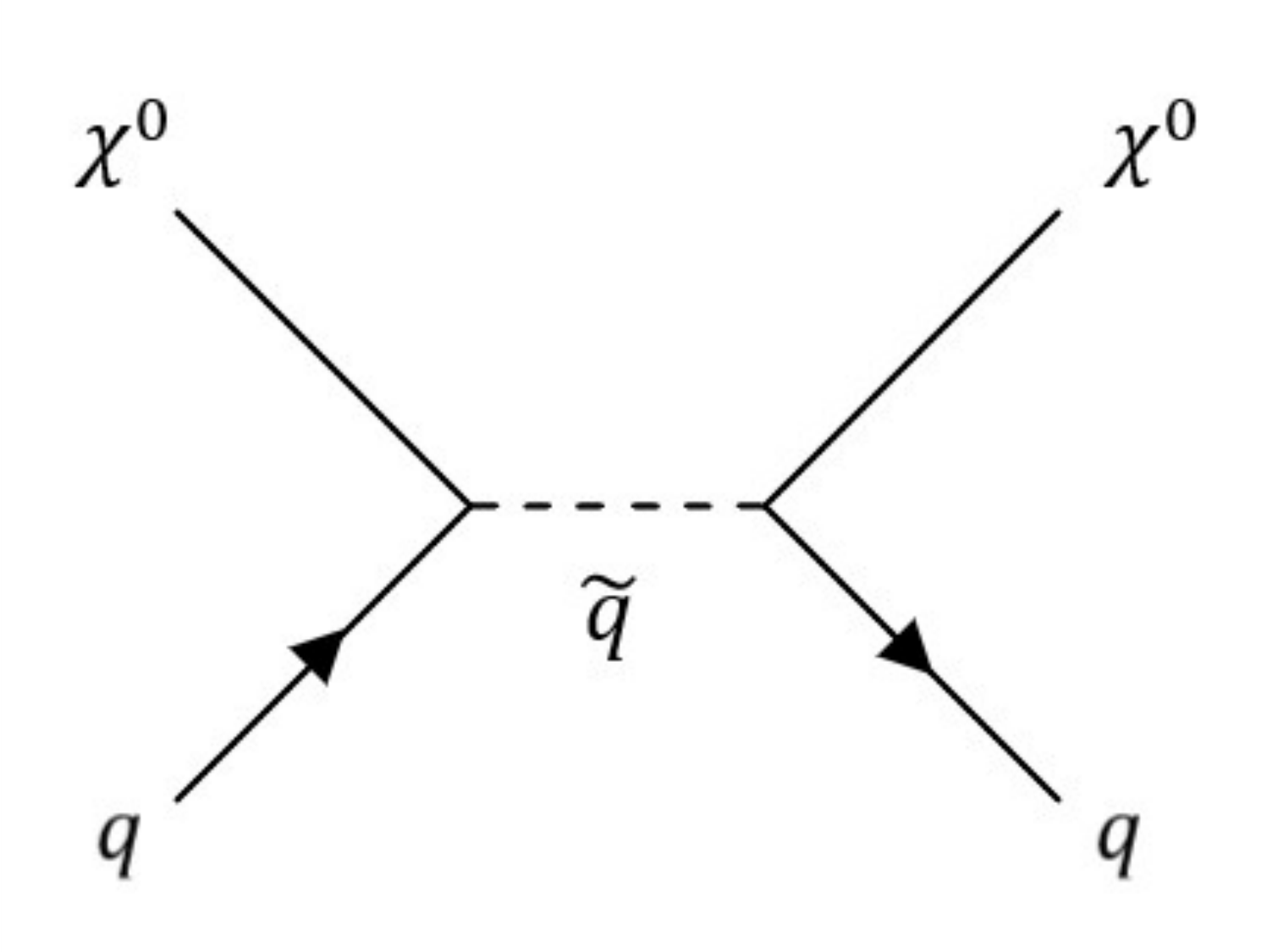}}
\resizebox{0.20\columnwidth}{!}{
\includegraphics{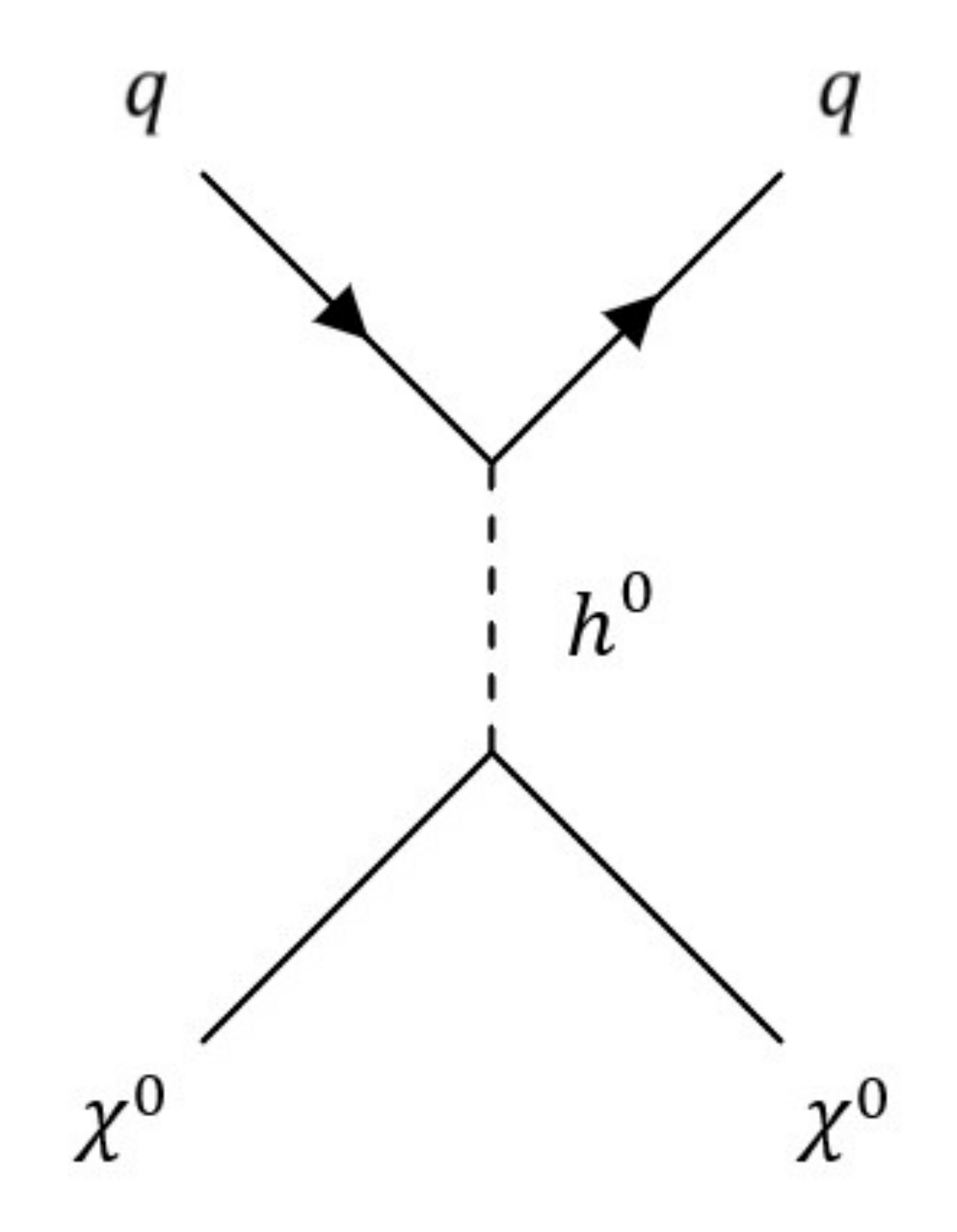}}
\resizebox{0.305\columnwidth}{!}{
\includegraphics{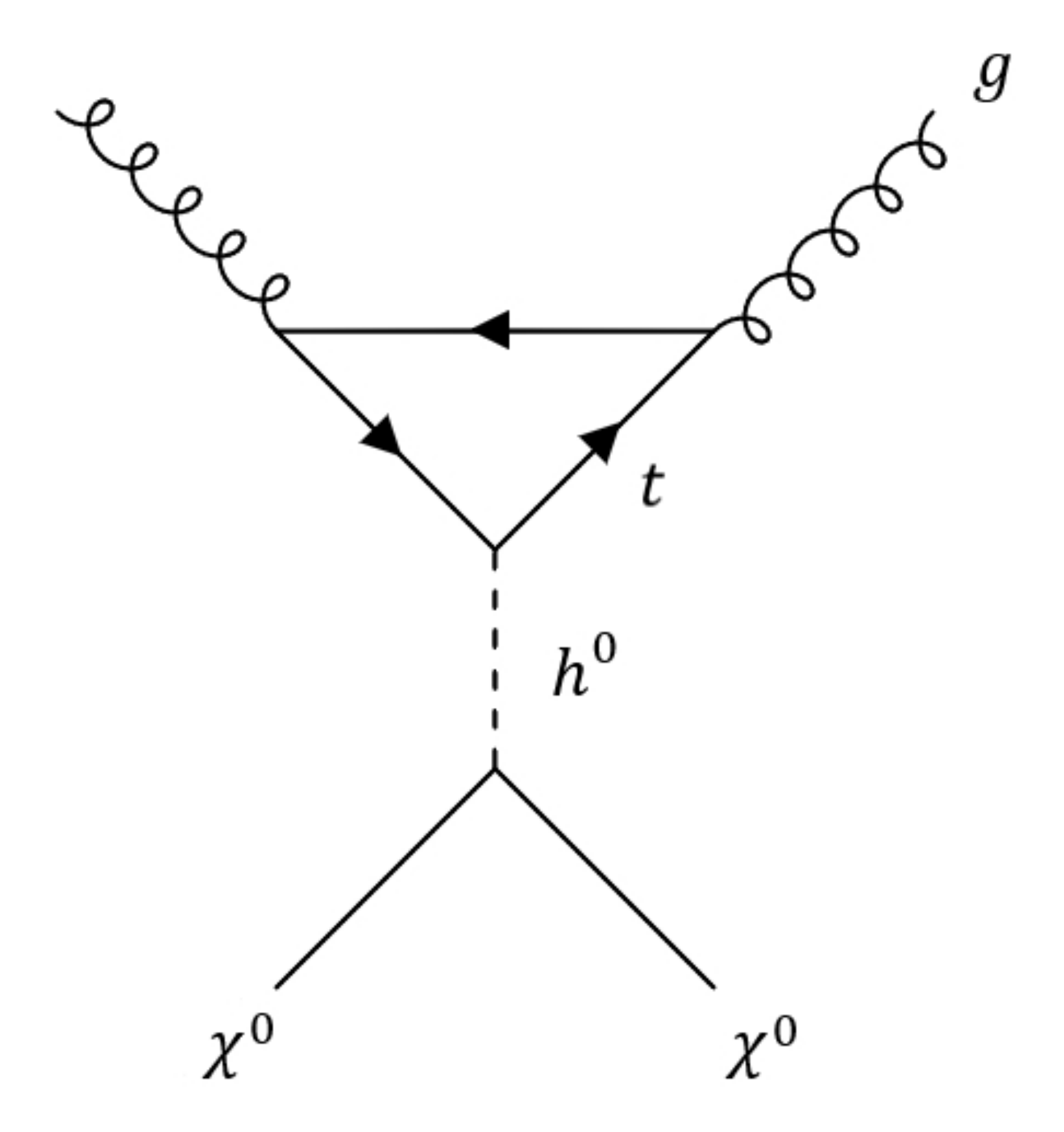}}
\end{center}
\caption{Left panels: Scattering off an atomic nucleus via $Z^0$ exchange and squark exchange occurs for neutralinos but not higgsons. 
Right panels: 
Scattering off an atomic nucleus via Higgs exchange leads to strong spin-independent scattering for natural neutralinos which are dominantly higgsino but have a significant gaugino admixture. }
\label{fig2}    
\end{figure}

In Figs.~\ref{fig1} and \ref{fig2} we show some of the processes involved in annihilation and scattering of neutralinos. These and many other processes are not available to higgsons, which can therefore have much smaller cross-sections. One of the most important aspects of the present theory is that the processes relevant to dark matter production and detection -- shown in Fig.~\ref{fig3} -- are very much simplified and weakened compared to those of susy models.

\section{\label{sec:sec3}Results}

\begin{figure}
\begin{center}
\resizebox{0.245\columnwidth}{!}{
\includegraphics{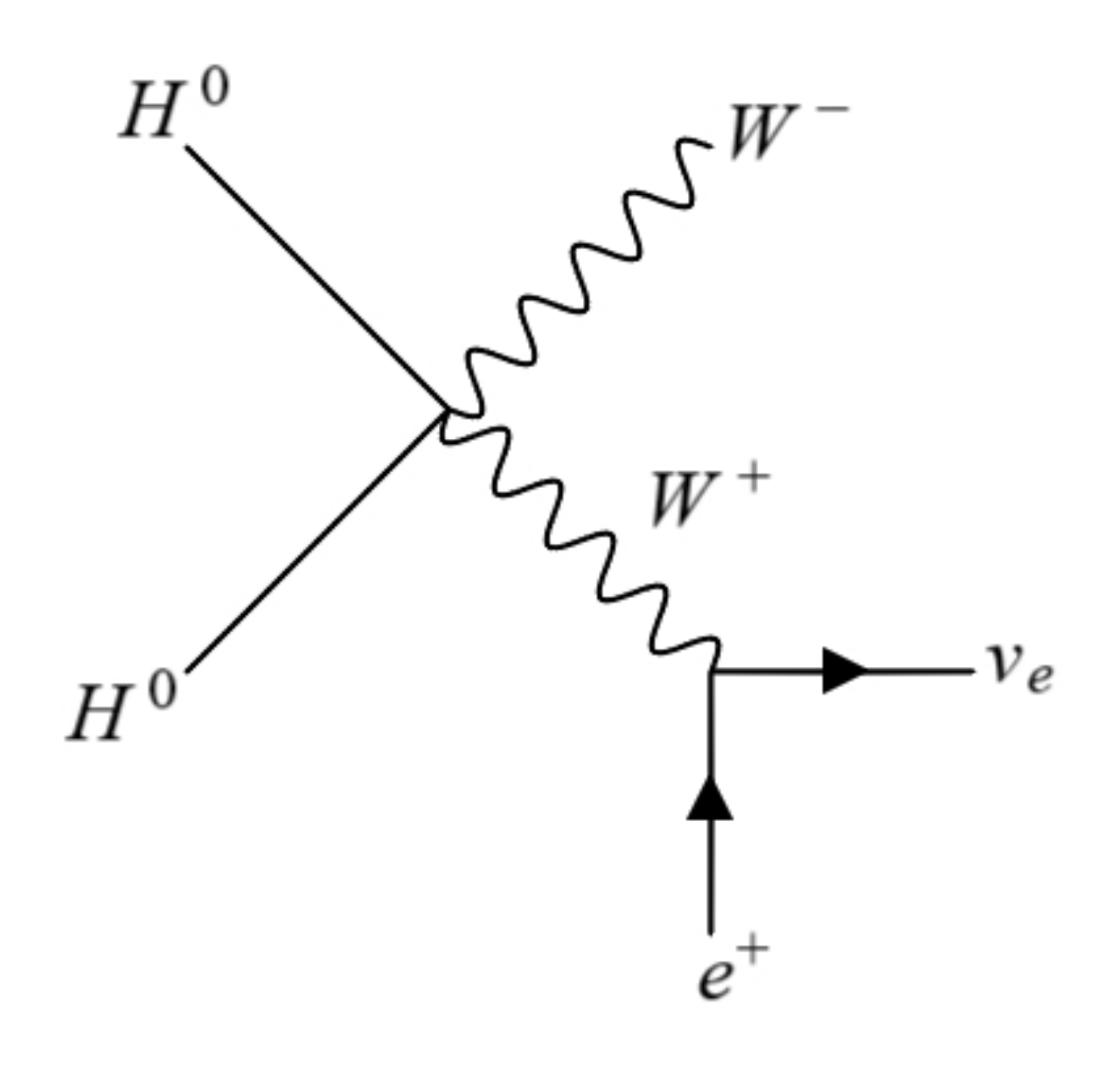}}
\resizebox{0.225\columnwidth}{!}{
\includegraphics{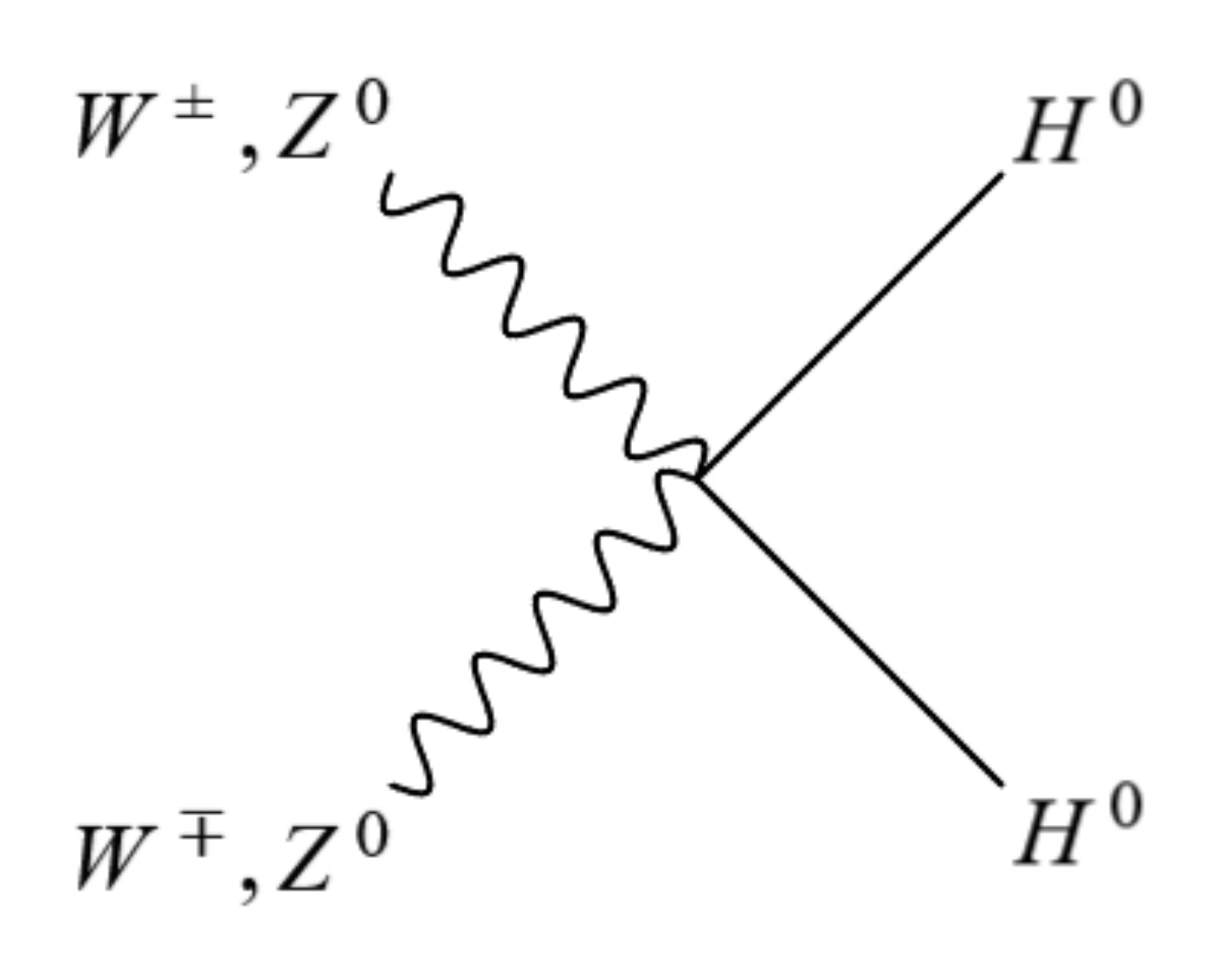}}
\resizebox{0.245\columnwidth}{!}{
\includegraphics{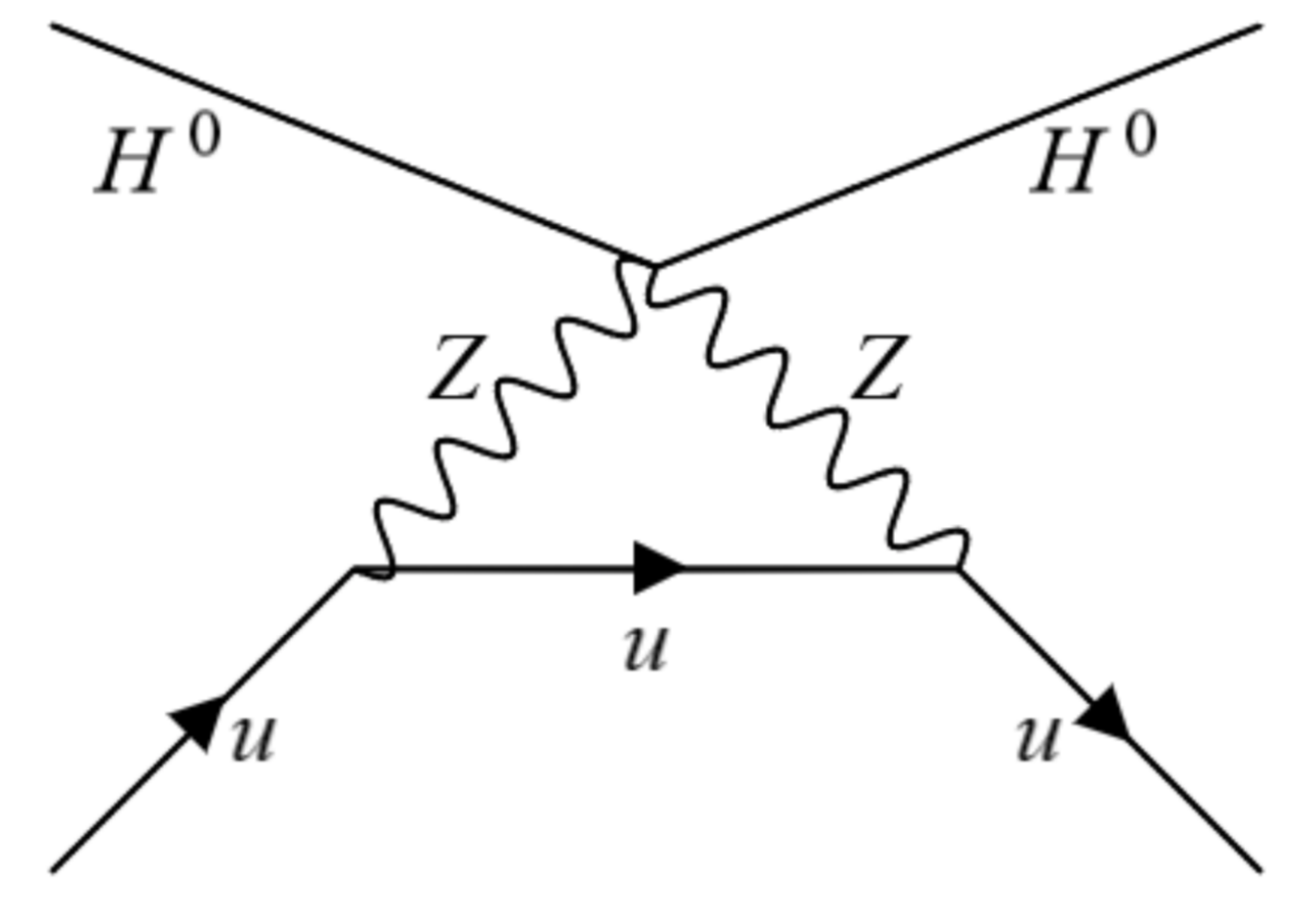}} 
\resizebox{0.245\columnwidth}{!}{
\includegraphics{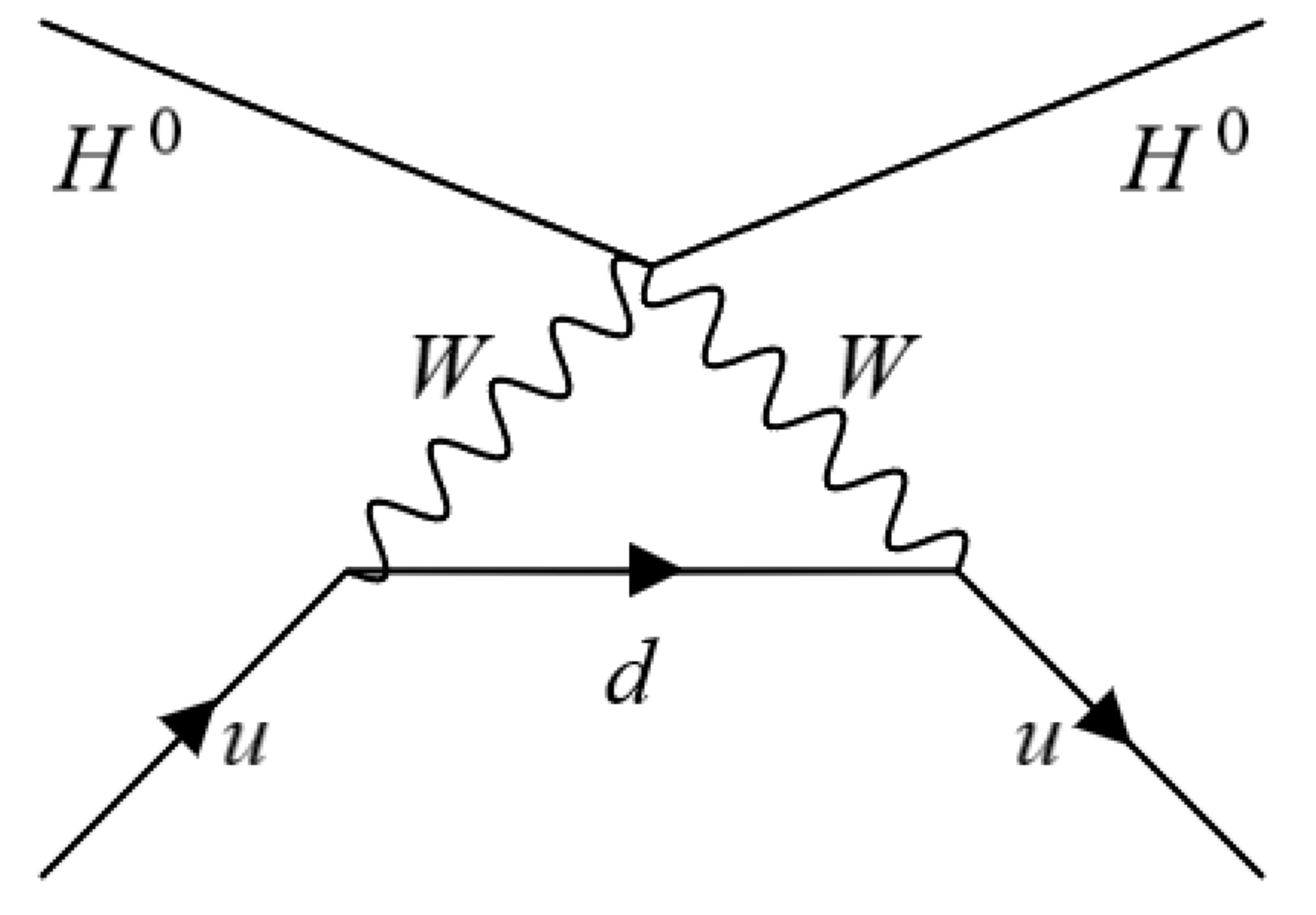}}
\end{center}
\caption{From left to right: (i) Example of annihilation into one real and one virtual
W boson; there are 18 such processes for
fermion-antifermion pairs (6 for leptons and 12 for the quarks with
sufficiently low masses). 
(ii) Collider creation through vector boson fusion.
(iii) and (iv) Scattering off a quark in an atomic nucleus through $Z$ and $W$ loop processes.}
\label{fig3}
\end{figure}

We have performed approximate calculations of the annihilation
cross-sections for processes shown in the left-most panel of Fig.~\ref{fig3}, using
standard methods~\cite{peskin,cheng}, and making the approximation of neglecting
the masses of the fermions (which are all small compared to $m_Z$, $m_W$,
and $m_H$). We find that 
\begin{align}
\langle \sigma _{ann}v\rangle \sim \langle \sigma v\rangle _{S}\quad 
\mathrm{with}\quad m_{H}\sim 72\,\mathrm{GeV}
\label{eq10}
\end{align}
where $\langle \sigma v \rangle_S = 2.2 \times 10^{-26} \, \mathrm{cm}^3/\mathrm{s}$
is the benchmark value obtained by Steigman et al.~\cite{Steigman} for a
WIMP with mass above 10 GeV that is its own antiparticle, if the relic dark
matter density is to agree with astronomical observations.

We have also made crude estimates 
of the scattering and creation cross-sections, but for these we principally rely on 
previous detailed calculations for closely related processes. In particular, the cross-sections for $H^0$ 
will be the same as those for  $H_{I}^{0}$ in the inert doublet model (IDM)~\cite{Eiteneuer}, which adds an extra field 
\begin{align}
\left( 
\begin{array}{c}
H_{I}^{+} \\ 
\frac{1}{\sqrt{2}} \left( H_{I}^{0}+ i A_{I}^{0} \right)
\end{array}
\right) \; ,
\label{eq0}
\end{align}
if (i) the coupling of $H_{I}^{0}$ to the Higgs is set equal to zero and (ii) the masses of 
the other particles $A_{I}^{0}$ and $H_{I}^{+}$ are set far above that of $H_{I}^{0}$.
In addition, the cross-section for the creation processes on the right of Fig.~\ref{fig3} 
are comparable to those for double-Higgs production via vector boson fusion. Based on 
the calculations discussed below~~\cite{Dutta,Dercks-2019,Bishara-2017,Dreyer-2020,Klasen-2013},
we estimate that the cross-section for collider detection in an LHC proton-proton collision is $\sim 10^{-3}$ pb 
and that the cross-section for direct detection in a Xe-based experiment is slightly below $10^{-11}$ pb.

The (multi-channel) annihilation cross-section of (\ref{eq10}) is consistent with the limits set by
observation of gamma-ray emissions from dwarf spheroidal galaxies by
Fermi-LAT~\cite{Leane-1,gammas-okay,Leane-2}, 

This cross-section and mass are also consistent with analyses of the gamma
ray excess from the Galactic center observed by Fermi-LAT~\cite{Goodenough,Fermi,Fermi-GCE,Leane,Cuoco2}, 
and with analyses of the antiprotons observed by AMS~\cite{Cuoco2,Cuoco,Cui,AMS-1,AMS-2}, 
which independently have been interpreted as potential evidence of dark
matter annihilation. The inferred values of the particle mass and
annihilation cross-section are in fact remarkably similar to those obtained
here; see e.g. the abstracts of Refs.~\cite{Fermi-GCE} and \cite{AMS-1}, and
Fig. 12 of Ref.~\cite{AMS-2}.

The predictions of the present theory within this context are also very similar to those for the IDM~\cite{Eiteneuer} if the 
masses of $A^0_I$ and $H^{\pm}_I$ are well separated from that of $H^0_I$ and the Higgs coupling is small.
A very detailed analysis of the Fermi-LAT
gamma-ray data, and its comparison with IDM predictions, has been given in Ref.~\cite{Eiteneuer}.
For annihilations of a dark matter WIMP with a mass of $\sim 72$ GeV  
the basic qualitative conclusions are the same as above.

The Planck observations of the CMB have ruled out a dark matter interpretation of the AMS positron excess~~\cite{Planck}, but are quite 
consistent with a dark matter particle having the mass and annihilation cross-section of the present candidate. Further clarification may come from future experiments such as the Cherenkov Telescope Array~\cite{CTA}.

The present particle is also consistent with the current 
collider-detection limits. The best possibility for creation in
a collider experiment appears to be the process depicted in Fig.~\ref{fig3}.
In the present context this is a weak process, but the results of
Refs.~\cite{Bishara-2017} and \cite{Dreyer-2020}  (for double Higgs production) 
and \cite{Dutta} and \cite{Dercks-2019} (for the IDM), 
indicate that the cross-section for LHC proton-proton scattering is of $\sim 1$ fb, and that observation of this process should ultimately be possible with a sustained
run of the high-luminosity LHC (if an integrated luminosity of up to 3000 fb$^{-1}$ can be achieved). 
Definitive collider studies may have to await a 100 TeV hadron collider or a very high energy lepton collider.

The best prospect for direct detection appears to be the one-loop processes shown in 
Fig.~\ref{fig3}, which are the same as for the IDM in Fig.~1 of  \cite{Klasen-2013}. 
The results in e.g. Figs. 2, 3, and 7 of that paper -- as well as the general conclusion ``$\sigma_{SI}$ (1-loop) is never below $\sim 10^{-11}$ pb'' indicate that this mechanism -- scattering via an exchange of two Z or W bosons -- has 
a cross-section slightly below $10^{-11}$ pb for the present particle, which should be  
within reach of the LZ and Xenon nT direct-detection experiments now under construction. It should be
well within reach of Darwin, the planned ultimate extension of the Xenon series of experiments, and certainly above the neutrino floor~\cite{Strigari-2014}.

\section{\label{sec:sec4}Conclusion}

The results described above -- from approximate calculations, estimates, and comparisons with calculations for closely related processes -- make it clear that the present dark matter candidate is consistent with current experiments and observations, while also potentially observable within the foreseeable future through a range of experiments. We plan to perform more detailed and precise calculations using the specialized software that has been developed for this purpose, such as MadDM~\cite{MadDM} for relic abundance, indirect detection, and direct detection, and Madgraph, Pythia, and Delphes for collider detection.

It should be mentioned that the lowest-mass excitation $H^0$ is not the only stable higgson, but it will emerge
from the early universe with the highest density: The more massive particles will fall out of equilibrium earlier and be
rapidly thinned out by the subsequent expansion, and they also will have larger annihilation cross-sections. The higher mass higgsons should eventually be observable in collider experiments at higher energies.

The fact that the present WIMP is fully consistent with experiment and observation, but still observable in the near future through multiple experiments, suggests that the patient search for WIMPs may soon be rewarded. Historic achievements have never been easy -- the Higgs boson, gravitational waves, black holes, and atoms were respectively observed about a half century, century, two centuries, and two millennia after they were predicted -- and it will not be surprising if WIMP detection comes only after decades of heroic efforts.

\end{document}